\documentclass[a4paper]{aa}
\usepackage{natbib}
\usepackage{epsfig}

\begin{document} 

\newcommand{\es}{erg s$^{-1}$}                          
\newcommand{\halpha}{H$\alpha$}                   
\newcommand{\hbeta}{H$\beta$}
\newcommand{\kms}{km~s$^{-1}$}       
\newcommand{\cmthree}{cm$^{-3}$}
\newcommand{\msun}{M$_{\odot}$} 

\title{Discovery of X-ray emission from the protostellar jet L1551 
IRS5 (HH 154)}

\author{F. Favata\inst{1} \and C.\,V.\,M. Fridlund\inst{1} \and
G. Micela\inst{2} \and S. Sciortino\inst{2} \and A.\,A. Kaas\inst{3}}

\institute{Astrophysics Division -- Space Science Department of ESA, ESTEC,
  Postbus 299, NL-2200 AG Noordwijk, The Netherlands
\and
Osservatorio Astronomico di Palermo, 
Piazza del Parlamento 1, I-90134 Palermo, Italy 
\and
Nordic Optical Telescope, Apartado 474, E-38700 Santa Cruz de la 
Palma, Canarias, Spain
}

\offprints{F. Favata,\\ Fabio.Favata@rssd.esa.int}

\date{Received date / Accepted date}


\abstract{ We have for the first time detected X-ray emission
  associated with a protostellar jet, on the jet emanating from the
  L1551 IRS5 protostar. The IRS5 protostar is hidden behind a very
  large absorbing column density, making the direct observation of the
  jet's emission possible. The observed X-ray emission is likely
  associated with the shock ``working surface'', i.e.\ the interface
  between the jet and the circumstellar medium. The X-ray luminosity
  emanating from the jet is, at $L_{\rm X} \simeq 3 \times 10^{29}$
  \es, a significant fraction of the luminosity normally associated
  with the coronal emission from young stars. The spectrum of the
  X-ray emission is compatible with thermal emission from a hot
  plasma, with a temperature of $\simeq 0.5$ MK, fully in line with
  the temperature expected (on the basis of the jet's velocity) for
  the shock front produced by the jet hitting the circumstellar
  medium.  \keywords{ISM: clouds -- ISM: individual objects: L1551 :
    HH 454-- ISM: jets and outflows -- Stars: formation -- Stars:
    pre-main sequence -- Radio lines: ISM } } \maketitle

\section{Introduction}
\label{sec:intro}

During the final stages of the formation of low-mass stars (in the
so-called classical T~Tau phase) accretion of material from the
proto-stellar nebula onto the Young Stellar Object (YSO) takes place
through an accretion disk. Very often (and possibly always) the
presence of the accretion disk is correlated with the presence of
energetic polar outflows, that is, collimated jets of material being
ejected perpendicularly to the disk, along its axis. Several models of
the formation of the jet have been proposed, in most of which the jet
is collimated by the presence of a (proto-stellar) magnetic
field. When these jets collide with the surrounding ambient medium --
or with previously ejected material -- they form a shock structure,
which is directly observable in the form of so-called Herbig-Haro jets
(e.g.\ \citealp{rr99}).

X-ray emission (and thus the presence of hot plasma, at temperatures
in excess of several $\times 10^5$ K, up to $\simeq 100$ MK during
energetic flares) has by now been observed in most stages of the
formation of low-mass stars, ranging from the highly embedded, perhaps
spherically accreting protostars (Class~0 objects) to the final stages
of the pre-main sequence life of a star, the Weak-Line T~Tau stage,
during which the X-ray luminosity is thought to come from a ``normal''
(however very active) stellar corona.

While accretion itself has been considered as a possible source of
X-ray emission in classical T~Tau stars, up to now no evidence of
energetic phenomena associated with protostellar jets has been
observed. In this paper we present the first observations of X-ray
emission from a protostellar jet, obtained in a well-studied system in
which the proto-star (and its immediate circumstellar environment)
powering the outflow is so heavily obscured that the jet can be
singled out as the source of emission of the X-rays without ambiguity.
Our observations show that this jet is indeed an X-ray source with a
luminosity equivalent to a fraction of the X-ray luminosity normally
associated with YSOs. The observed X-ray spectrum is compatible with a
thermal origin of the observed X-ray emission. The associated
temperature is moderate, well matched to the shock velocities observed
in this and other Herbig-Haro jets. This raises the question of
whether the X-ray emission associated with jets could indeed be a
common feature of stellar formation, so that in some cases a
significant fraction of the X-ray luminosity associated with the star
(YSO/accretion disk) is actually emanating from shocks in the jet.

\section{The L1551 IRS5 outflow}
\label{sec:sample}

The L1551 cloud is one of the nearest ($d \simeq 140$ pc) sites of
ongoing star formation, in which objects in several different stages
of the process are clearly visible, from deeply embedded, actively
accreting (proto-)stars to the final stages of star formation
represented by the Weak-Line T~Tau stars with no remaining
circumstellar material. Several ``canonical'' examples of jets and
outflows associated with protostellar accretion are present in the
region.  In this paper we are mainly concerned with the jet associated
with the IRS5 source embedded in the L1551 cloud and its associated
outflow. 


L1551 IRS5 is a deeply embedded protostellar binary system (e.g.\ 
\citealp{rod98} and references therein), effectively invisible at
optical wavelengths as it is hidden behind some $\simeq 150$ mag of
visual extinction (\citealp{shs+88}; \citealp{smi87};
\citealp{wlm+2000}) which most likely originates in the circumstellar
accretion disk. The two Class 0/1 stars have a total luminosity of
$\simeq 30~L_\odot$. The two IRS5 stars appear to be (jointly?)
powering at least two observable outflows. A large (several arcmin)
bipolar molecular outflow (actually the first discovered,
\citealp{sne80}) and a much smaller (with a length of $\simeq 10$
arcsec) denser two-component jet (\citealp{fl98}), consisting of
material at temperatures of $T \simeq 10^4$ K, thus visible in the
emission lines of e.g.\ H$\alpha$, representative of recombination.
The jet and the molecular outflow have been shown to be likely
causally unrelated, given that the jet has a momentum insufficient by
several orders of magnitude to drive the molecular outflow
(\citealp{fl98}). The jet moves at transverse velocity of 200--400
\kms\ (\citealp{fl94}) and appears to end in a shock against the
ambient medium (a ``working surface'' -- \citealp{fl98})

\section{Observations}
\label{sec:obs}

\subsection{XMM-Newton}

The X-ray observations discussed in the this paper were obtained with
the XMM-Newton observatory. A deep (50 ks) exposure of the
star-forming region of the L1551 cloud was obtained starting on Sep.\ 
9 2000 at 19:10 UTC. All three EPIC cameras were active at the time of
the observation, in full-frame mode, with the medium filters.

Data have been processed by us with the standard SAS V5.0.1 pipeline
system, concentrating, for the spectral analysis, on the EPIC-PN
camera. In order to minimize the unwanted contribution of non-X-ray
events we have retained only the counts whose energy is in the
0.3--7.9 keV range. To deal with the time-varying background, we have
applied a technique purposely developed at the Palermo Astronomical
Observatory which maximizes the statistical significance of weak
sources by identifying and removing the fraction of the exposure time
strongly affected by high-background episodes.

\begin{figure}[!tbp]
  \begin{center} \leavevmode \epsfig{file=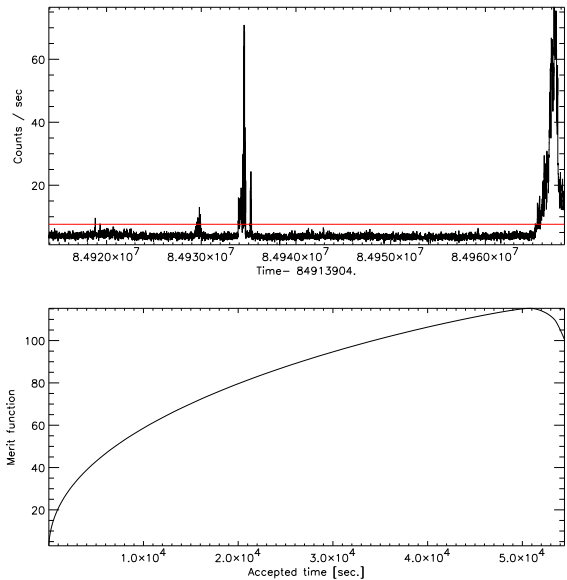, height=8.0cm,
      angle=0}
  \caption{The top panel shows the total count rate as a function of
    time in the PN camera during the complete L1551 XMM observation
    discussed in the present paper. The horizontal line is the chosen
    threshold value for background filtering: time intervals during
    which the total PN count rate was above the line have been
    discarded from the final data set. The bottom panel shows the
    value of the merit function, as a function of the total accepted
    observing time, which was used to determine the threshold. The
    merit function is optimized to yield an optimal data set for the
    detection of faint sources.}
  \label{fig:bck}
  \end{center}
\end{figure}

In many XMM data sets most of the background is due to a small number
of short (but intense) episodes -- mostly related to solar events --
so that the background events are strongly concentrated in time.
Removal of the small fraction of the observing time in which the
high-background episodes are concentrated results in a much larger
$S/N$ ratio data set with a comparatively small loss of source
photons.  Fig.~\ref{fig:bck} shows the time evolution of the total
count rate for the PN data set (in the top panel), together with the
threshold (shown by the horizontal line) above which the data are
discarded.  The threshold is dynamically chosen by maximizing a merit
function, defined to optimize the detection of faint sources (which
are background, rather than photon-noise, dominated). The value of the
merit function for the L1551 PN observation is plotted in the lower
panel of Fig.~\ref{fig:bck} as a function of the total accepted time.

As a result of the overall procedure the initial data set of $\simeq
10^6$ photons collected during an exposure of $\simeq 56.8$ ks for the
MOS cameras and $\simeq 54.5$ ks for the PN camera, was reduced to a
cleaned data set of $\simeq 250\,000$ photons collected in $\simeq 55$
ks for the MOS and $\simeq 51$ ks for the PN cameras, i.e with a
judicious time and energy filtering we have been able to reduce the
background level by a factor $\simeq 4$ while rejecting only $\simeq
5$\% of the overall exposure time.

The cleaned data set obtained by summing the data of the two MOS and
one PN EPIC cameras has been searched for sources with the Wavelet
Transform detection algorithm also developed at Palermo Astronomical
Observatory (\textsc{Pwdetect}, Damiani et al.\ 2001 in preparation).
The characteristics of the XMM-Newton version of the algorithm are
inherited from the version developed for the ROSAT PSPC
(\citealp{dmm+97a}). The overall analysis procedure follows the recipe
described in more details by \citet{smd+2001}. The L1551 observation
has been taken with the medium filters; in such a case we have derived
that the value of the relative efficiency of the PN and of the
individual MOS cameras is 2.94, hence the summed data set has a single
MOS-equivalent cleaned exposure time of $2 \times 55 + 2.94 \times 51
= 260 $ ks.

In order to assess the source significance threshold to adopt to
ensure detection of the faintest sources a full set of simulations of
empty fields with the same background level and the same exposure map
as the summed data set needs to be run. Following this, a limiting
threshold can be set which ensures e.g.\ a maximum of one spurious
source per field. While this has been done for the complete analysis
of the full L1551 exposure (and will be discussed in a future paper),
the three X-ray sources discussed in the present paper are all well
above the minimum threshold, and consequently their significance is
well above the spurious source threshold.


\begin{figure}[!tbp]
  \begin{center} \leavevmode \epsfig{file=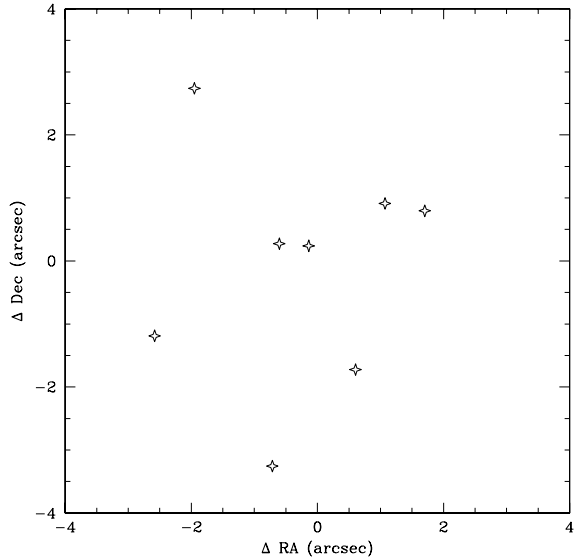, height=8.0cm,
      angle=0}
  \caption{The plot shows the difference (in arcsec) between the
    coordinate of the X-ray source as determined in the PN detector by
    the \textsc{Pwdetect} algorithm and their optical counterpart, for
    the 8 bright X-ray sources (with more than 300 counts) with
    unambiguous counterparts in the L1551 XMM observation.}
  \label{fig:deltac}
  \end{center}
\end{figure}

The quality of the XMM aspect solution (``bore-sight correction'') was
verified by comparing the positions of X-ray bright sources as
determined in the PN detector by the \textsc{Pwdetect} algorithm with
the position of their optical counterparts. Only X-ray bright sources
(with more than 300 source counts in the PN camera) with single,
unambiguous optical counterparts with good position determination were
used, resulting in 8 such sources being available in the L1551 field.
The optical coordinates were obtained from the \textsc{Simbad}
database. No significant bias is present in the data, with a mean
$\Delta {\rm RA} = -0.33 \pm 1.46$ arcsec, and a mean $\Delta {\rm
  Dec} = -0.15 \pm 1.85$ arcsec, so that the positional error for the
faint sources is fully dominated by the photon noise coupled with the
XMM point spread function.

The extraction of source and background photons was performed using
the \textsc{xmmselect} tool. Source photons have been extracted from a
circular region of 45 arcsec diameter, while background photons have
been extracted from a region on the same CCD chip and at the same
off-axis angle as for the source region.  Response matrices
(``\textsc{arf} files'') appropriate for the position and size of the
source extraction regions have been computed for the EPIC PN camera.
The spectral analysis has been performed using the XSPEC package,
after rebinning the source spectra to a minimum of 20 source counts
per (variable width) bin.

\begin{figure*}[tbp]
  \begin{center} \leavevmode \epsfig{file=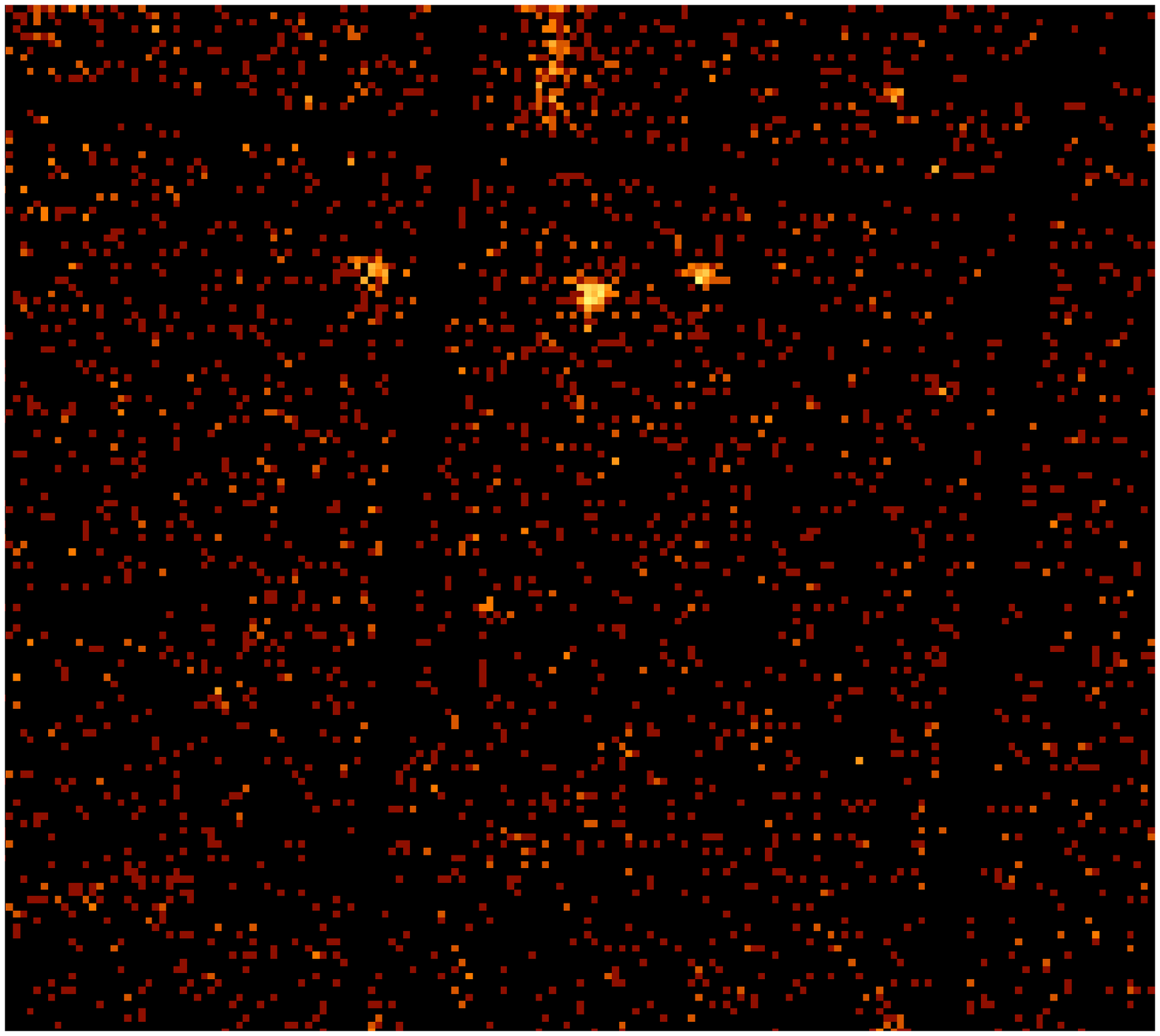, width=8.0cm,
      angle=0} \epsfig{file=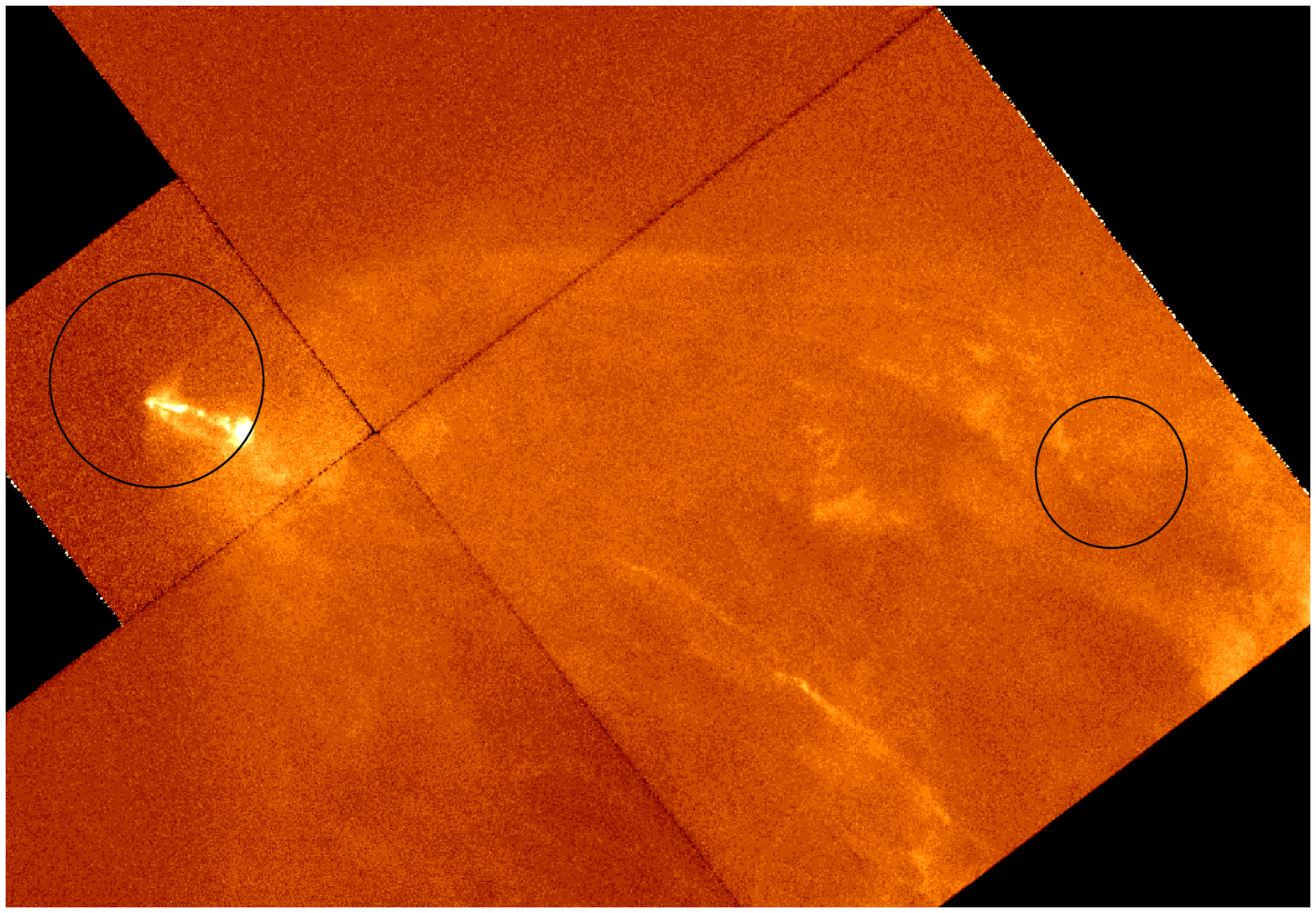,  width=8.0cm,
      angle=0}
  \caption{The left panel shows the region of L1551 IRS5 in X-rays, as
    seen in the XMM EPIC-PN camera, while the right panel shows a
    small part of the same region as seen in a 1800 s $R$-band CCD
    image obtained with the Hubble Space Telescope WFPC2 camera. The
    size of the small detector (the WFPC2-PC chip) on the left part of
    the HST image is 36.8 arcsec, while the size of the X-ray image is
    $9.3 \times 6.3$ arcmin.  The position of the two leftmost X-ray
    sources visible in the left panel is indicated on the $R$-band
    image by the circles.  The leftmost X-ray point source is the one
    associated with L1551 IRS5, while the other point source is one of
    the two background sources discussed in Sect.~\ref{sec:bck}. The
    nearly vertical alignment of bright pixels near the center-top of
    the X-ray image is due to ``spill-over'' from a bright source
    higher up in the image.}
  \label{fig:image}
  \end{center}
\end{figure*}

\subsection{Ground-based observations}

New optical spectra and images of the IRS5 jet have been obtained with
the 2.6\,m Nordic Optical Telescope (NOT). The spectra were obtained
in December 1999 using the echelle mode of the ALFOSC combined imager
and spectrograph.  The cross disperser/grating combination allowed for
a spectral resolution of $\simeq 1$ \AA\ at the wavelength of \halpha.
The detector was a $2048 \times 2048~ 15~ \mu$m pixel CCD. A slit
width of 1 arcsec projected on the sky was used, and the integration
time per spectrum was 3600 s. The complete velocity field of the jet
was mapped in emission lines between 4800 \AA\ and 7900 \AA. For the
purpose of evaluating shock velocities in this paper our
interpretation is based on an analysis of the \halpha\ data with the
other lines used as a consistency check.  A full analysis will be
presented in a forthcoming paper. \hbeta\ was used together with
\halpha\ in order to estimate the extinction towards the shocks in the
L1551 IRS5 jet.

For the identification of the optical counterparts to the X-ray
sources detected in the XMM image both new narrow band images (also
obtained with the NOT) and already available HST images were used.
The HST $R$-band WFPC2 image of the jet region is shown in
Fig.~\ref{fig:image} together with the EPIC PN X-ray image; the
position of the X-ray source associated with the L1551 jet, together
with the position of one of the background sources, is shown on the
WFPC2 image by a black circle. The size of the circle corresponds to
the likely extent of the X-ray source as determined by the
\textsc{Pwdetect} algorithm. The WFPC2 image was obtained in February
1996 with the F675W ($R$-band) filter, and part of it is discussed in
\citet{fl98}.

New narrow band images were also obtained with the NOT. These
observations were carried out in March of 2001 again using the ALFOSC,
this time in imaging mode. Deep images integrated for 1800 s (\halpha)
and 3600 s (I-band) were obtained in order to detect possible new
emission knots, as well as to search for faint background or embedded
sources shining through the molecular cloud (in possible association
with the background X-ray sources discussed in Sect.~\ref{sec:bck}). A
narrow band filter centered on the \halpha\ line was used, while the
I-band filter was centered on 7150 \AA\ with a FWHM of 1600 \AA.
Being several arcmin in size the field of view of the NOT images is
significantly larger than the WFPC2 one, although with less spatial
resolution and shallower limiting magnitude.

\section{X-ray emission from the L1551 IRS5 jet}

\begin{figure}[tbp]
  \begin{center} \leavevmode \epsfig{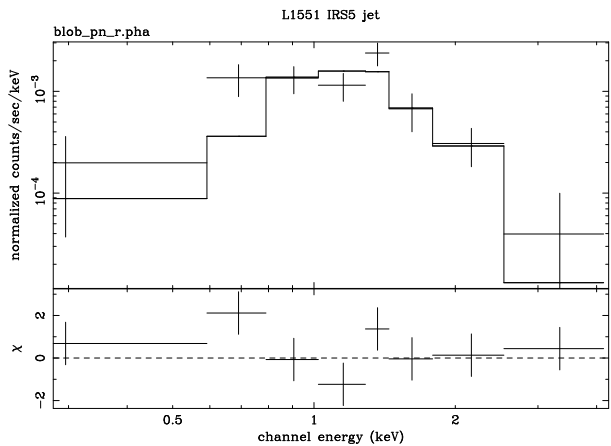}
  \caption{The observed, background-subtracted EPIC PN X-ray spectrum
    of the X-ray source associated with the L1551 IRS5 jet. The
    best-fit thermal (\textsc{``mekal''}) spectrum -- whose parameters
    are given in Table~\ref{tab:source} -- is also shown.}
  \label{fig:irs5spec}
  \end{center}
\end{figure}

A faint X-ray source (the leftmost one in the left panel of
Fig.~\ref{fig:image}) is positionally coincident with the embedded
source L1551 IRS5 and its jet. The background-subtracted count rate is
only $8.4 \times 10^{-4}$ cts~s$^{-1}$ in the EPIC PN camera, so that
the total number of source counts is limited to 42 cts in the $\simeq
50$ ks exposure (with an equivalent number of background counts in the
extraction box used, a circle 45 arcsec in diameter). The low
statistics allow a limited amount of spectral information to be
derived for the source. The resulting spectrum (shown in
Fig.~\ref{fig:irs5spec}) is soft, and can be reasonably described with
a moderately absorbed thermal spectrum. Using a \textsc{mekal} model
with an added interstellar absorption component in \textsc{xspec}
gives a best-fit temperature $T = 0.5 \pm 0.3 \times 10^6$~K, with a
moderate value of the best-fit absorption ($1.4 \pm 0.4 \times
10^{22}$ cm$^{-2}$), corresponding to an extinction of $A_V = 7.3 \pm
2.1$ mag\footnote{Using a conversion factor $N_{\rm H}/A_V = 1.9
  \times 10^{21}$ atoms cm$^{-2}$ mag$^{-1}$, see e.g.\ 
  \citet{cox2000}}. The null hypothesis probability for the fit is
$15\%$, i.e.\ the fit is, from the statistical point of view, fully
acceptable. The limited statistics of the spectrum (and the
corresponding small number of bins) do not allow to constrain the
spectrum at a more detailed level. The full width at half power (FWHP)
of the XMM point-spread function (PSF) for EPIC PN camera is $\simeq
14$ arcsec, significantly larger than the size of the jets (whose
visible length is $\simeq 10$ arcsec).  Thus, it is not possible to
locate the precise site of the X-ray emission within the jet
structure.

\section{Background sources}
\label{sec:bck}

Two additional X-ray source are detected within the area on the sky
that is dominated by the molecular outflow from L1551 IRS5. The
centroid positions, and background subtracted count-rates can be found
in Table~\ref{tab:source}. The position of the brighter of the two
against the molecular outflow is visibile in Fig.~\ref{fig:image}.
Neither of these sources can be associated with any visible object in
the deep \halpha\ and $I$-band images at these positions.

\begin{figure}[tbp]
  \begin{center} \leavevmode \epsfig{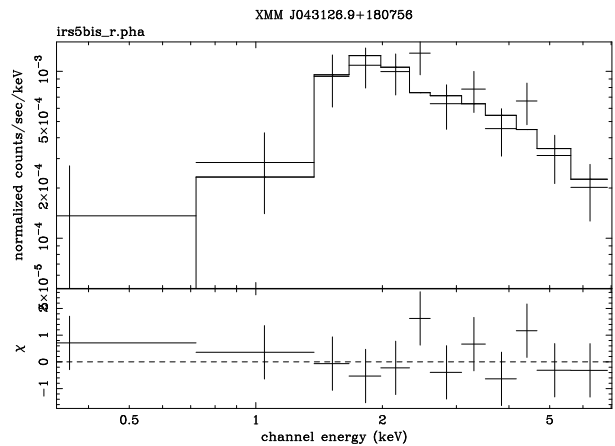}
  \caption{The observed, background-subtracted EPIC PN X-ray spectrum
    of the X-ray source XMM J$043126.9+180756$. The best-fit absorbed
    power-law spectrum -- whose parameters are given in
    Table~\ref{tab:source} -- is also shown.}
  \label{fig:irs5bisspec}
  \end{center}
\end{figure}

Both $\chi^2$ and K-S tests show that the X-ray emission from these
sources is constant to a high ($\ge 90\%$) probability level.  The
spectra of these two sources are significantly harder than the
spectrum of the source associated with the L1551 IRS5 jet (the
spectrum of the brighter of the two is shown in
Fig.~\ref{fig:irs5bisspec}), and can both be satisfactorily described
with an absorbed power-law spectrum, with indices varying between 1.2
and 2.5. The absorbing column density is in both cases moderate, with
$A_V$ between $\simeq 5$ and $\simeq 8$ magnitudes.  The resulting
column densities are thus similar to what is expected for the L1551
molecular cloud at these positions (\citealp{sb80}). Given the X-ray
spectral characteristics, as well as the lack of any visible candidate
counterpart in our deep $I$-band images, we consider it likely that
these sources are not associated with the molecular L1551 IRS5
outflow, even though they are positionally coincident with (parts of)
it. Rather, they are most likely to be extra-galactic X-ray sources
(plausibly active galactic nuclei) shining through the L1551 cloud.
Typical active galactic nuclei would have, on the basis of their high
$F_{\rm X}/F_V$ ratio, optical magnitudes $V \ga 19$, which would
become $V \ga 24$ when the intervening column density is taken into
account. Therefore their optical counterparts are not expected to be
visible, on the optical images presented here, againts the background
of the molecular outflow emission.


The spectra could in principle be fit also with a thermal spectrum,
with a very high resulting temperature ($T \ga 100$ MK). While this
temperature would not be incompatible with coronal protostellar origin
during e.g.\ an energetic flare, the lack of any visible counterpart
down to faint magnitudes (even though the absorbing column is only a
few magnitudes) and the constant light curve of the sources make this
last hypothesis unlikely.

\begin{table*}[thbp]
  \begin{center}
    \caption{Coordinates, count rates and best-fit spectral parameters
      for the three X-ray sources discussed in the present paper. The
      first source is the one associated with the L1551 IRS5 jets,
      while the two other ones are the sources positionally associated
      with the L1551 IRS5 molecular outflow but most likely not
      physically associated with it. The count rate is given in units
      of counts per ks, while the column ``$\sigma$'' contains the
      significance of the source detection in the combined MOS+PN
      image, in units of ``equivalent $\sigma$'' (i.e.\ obtained by
      integration of a Gaussian probability distribution). The
      absorbing column density is given in units of $10^{22}$
      cm$^{-2}$. For the X-ray source associated with the jets the
      best fit temperature in MK is given, while for the two sources
      not associated with the outflows the best-fit power-law index is
      given.  The last column gives the (unabsorbed) X-ray flux in the
      0.3--5.0 keV band in units of $10^{-13}$ erg cm$^{-2}$
      s$^{-1}$.} \leavevmode
    \begin{tabular}{lrrrrrrr}

Source     & $\alpha$ & $\delta$ & rate & $\sigma$ & $N({\rm H})$ & $T$ &
$F_{\rm X}$\\\hline
XMM J$043134.0+180806$   & 04 31 34.0 & 18 08 06 & $0.8\pm 0.3$ & 18.0
& $1.4\pm 0.4$ & $0.5 \pm 0.3$ & 1.3\\\hline
Source     & $\alpha$ & $\delta$ & rate  & $\sigma$ & $N({\rm H})$ & $\gamma$ &
$F_{\rm X}$\\\hline
XMM J$043126.9+180756$   & 04 31 26.9 & 18 07 56 & $3.4\pm 0.3$ & 24.6 & $2.3\pm 0.7$ & $2.0 \pm 0.5$ & 0.7\\
XMM J$043123.4+180805$   & 04 31 23.4 & 18 08 05 & $3.0\pm 1.4$ & 12.2
& $1.0\pm 0.3$& $2.6 \pm 1.0$ & 0.6\\
    \end{tabular}
    \label{tab:source}
  \end{center}
\end{table*}

The presence of two serendipitous X-ray sources in the PN image shown
in Fig.~\ref{fig:image} at flux levels of order $10^{-14}$ erg
cm$^{-2}$ s$^{-1}$ is fully in line with the expected number density
of background sources determined on the basis of the $\log N \log S$
relationship for X-ray sources (see e.g.\ \citealp{haa+2001}), which
predicts that at this flux limit 100 to 200 sources per square degree
should be present in any given X-ray observation. The area covered by
X-ray image of Fig.~\ref{fig:image} is $\simeq 0.02$ square deg, so
that the expected number of serendipitous sources is 2 to 4.

\section{Discussion}
\label{sec:disc}

Hubble Space Telescope (HST) observations (\citealp{fl98}) indicate
the presence of a number of shocks along the extent of the IRS5 jet.
The jet is observed to end in a ``working surface'' against the
ambient medium at $\simeq 10$ arcsec from the presumed location of
the source powering it (see Figs.\ 1 and 2 of \citealp{fl98}). This
shock feature is designated ``knot D'' in the nomenclature of
\citet{nec87} and \citet{fl94}.  We have measured the \halpha/\hbeta\ 
ratio towards the working surface (knot D) of the jet.  Assuming a
type B pure recombination spectrum (which is justified since we also
detect [O\,{\sc iii}] 5007 \AA\ emission at this position -- see
below), we find $A_V$ to be 4--6 mag, depending on which extinction
law is applied. We also find that the extinction is increasing in the
direction towards IRS5 along the jet (confirming the result of
\citealp{shs+88}).  The absorbing column density, for the IRS5 X-ray
spectrum, is thus compatible with the absorbing column density
measured toward the visible jet (and in particular toward the shock
feature), making the association between the X-ray emission and the
jet plausible. Since as mentioned above the IRS5 protostellar system
is hidden behind a very thick layer of absorbing material,
corresponding to $A_V \ga 150$ mag, it can be excluded that the X-ray
photons -- given the small absorbing column density and the lack of
high-energy photons in the spectrum) -- emanate from (or close to) the
photosphere/chromosphere of the protostars powering the jet.  We
therefore draw the conclusion that this source is the result of
thermal emission in the shocks whose recombination light is seen along
the jet in the visual wavelength regime.
 
The brightest shock as observed in several spectral lines at visual
wavelengths (\halpha, \hbeta, [O\,{\sc iii}], [O\,{\sc i}] and
[S\,{\sc ii}]) is the working surface (knot D -- see above). This is
also the only position along the jet where [O\,{\sc iii}] 5007 \AA\ 
emission has been unambiguously detected (\citealp{cf85};
\citealp{sne85}). This is thus indicative of shock velocities $\ga
100$ \kms\ (\citealp{hrh87}).  At this position the FWZI (Full Width
at Zero Intensity) of the \halpha\ emission line is 215 \kms, while
the peak radial velocity is $-180$ \kms. Taken together with the
proper motion of this knot of $\simeq 200$ \kms\ (\citealp{fl94})
this is indicative of a fluid velocity of $\simeq 270$ \kms, and an
inclination, $i \simeq 45$ deg. Following e.g.\ \citet{hrh87} we will
assume that the FWZI of the radial velocity is equivalent to the shock
velocity, which is also compatible with the presence of relatively
strong [O\,{\sc iii}] emission from this point. From the result of
\citet{fl98} we also know that the shock interface is extremely narrow
-- less than the spatial resolution of HST, which is estimated to be
$\leq 1.0 \times 10^{14}$ cm. The immediate post-shock temperature can
be estimated as in \citet{raga89}:

\begin{equation}
  T_{\rm ps} \simeq \frac{2.9 \times 10^5 {\mathrm K}}{1 + X} \times 
(\frac{v_{\rm shock}}{100\, {\rm km} {\rm s}^{-1}})^2
\end{equation}

where $X$ is the hydrogen pre-ionization fraction and a number
abundance of 0.9 (H) and 0.1 (He) has been assumed.  \citet{fl98} have
found that it is likely that the ionization fraction is close to 1.
Then, for a shock velocity of 215 \kms\ the resulting temperature will
be $T_{\rm ps} \simeq 0.67$ MK, fully compatible with the observed
X-ray temperature of $0.5 \pm 0.3$ MK. Thus, the observed X-ray
emission is likely to be due to material heated at the interface shock
(the working surface) between the jet and the ambient medium medium,
or possibly in shocks along the cavity excavated by the jet.

The X-ray luminosity of the emission associated with the jets is
$L_{\rm X} \simeq 3 \times 10^{29}$ \es (assuming a distance of 140 pc
for the L1551 complex). This value is approximately an order of
magnitude higher than the \halpha-luminosity of knot D, which is
$\simeq 4 \times 10^{28}$ \es (\citealp{fl94}).

\subsection{Energetics}

Following \citet{fl98}, we can attempt to determine the mass and
energetics of the jet, and compare these with the X-ray result. From
the [S\,{\sc ii}] 6717/6731~\AA\ ratio we find that the electron
density is $n_e \simeq 500$ \cmthree\ along the jet, while knot D --
identified as the working surface by \citet{fl98} -- has an electron
density of $n_e \simeq 10^4$ \cmthree. The ionization fraction in the
jet is close to 1 from the reasoning of \citet{fl98}. This is also
supported by the relatively low shock temperature found from the X-ray
spectrum.

We assume that the jet consists of the following components: a) a
``pipe flow'' of $1 \times 1 \times 10$ arcsec geometrical size and
density $n_e \simeq 500$ cm$^{-3}$ and b) a ``working surface'' of $2
\times 2 \times 2$ arcsec geometrical size and density $n_e \simeq
10^4$ cm$^{-3}$. Under these assumptions the mass of the jet is then
found to be between $1 \times 10^{-6}$~\msun\ and $2 \times
10^{-6}$~\msun\ (see also the detailed discussion in \citealp{fl98}).
The shock velocity is $\simeq 200$ \kms\ from the arguments above, and
the highest and lowest fluid velocities (measured proper motions and
$v_{\rm rad}$ corrected for inclination) are $\simeq 100$ \kms\ and
$\simeq 300$ \kms\ respectively. This means that the mechanical
luminosity of the jet will be between $10^{41}$ and $10^{42}$ erg
s$^{-1}$, so that a very low conversion efficiency between mechanical
and radiant luminosity is sufficient to justify the observed X-ray
luminosity from the shock. Again, the \halpha\ luminosity is $\simeq
4.7 \times 10^{28}$ \es\ (\citealp{fl98}, \citealp{fl94}),
approximately an order of magnitude lower the X-ray luminosity derived
here.

A characteristic size for the X-ray emitting region can be derived
from the emission measure determined from the X-ray spectrum and the
density determined above. The emission measure is defined as $E\!M =
\int n_e n_H dV \simeq 0.8\, n^2 V$, where $n_e$ is the electron
density, $n_H$ the hydrogen density, and $V$ is the volume of the
emitting region (under the simplistic assumption of uniform density).
Given that $E\!M = 1.1 \times 10^{52}$ cm$^{-3}$ (from the thermal fit
to the X-ray data), assuming a density $n_e \simeq 10^4$ \cmthree\ 
(see above), one derives a volume $V = 1.4 \times 10^{44}$ cm$^3$,
which corresponds to a characteristic linear size $l \simeq V^{1/3}
\simeq 5 \times 10^{14}$ cm. This scale size is comparable to the
upper limit for the size of the shock interface derived from the HST
observations, $\leq 1.0 \times 10^{14}$ cm, thus further supporting
the identification of the shock interface as the seat of the X-ray
emission.

\section{Conclusions}

While the energetic nature of the collimated jets observed to be
originating from protostellar sources has been evident for some time,
no high-energy photons have up to now been observed from these
phenomena. Here we report the first convincing evidence of X-ray
emission from the protostellar jet associated with the IRS5
protostar(s) in the L1551 cloud. The X-ray source and the protostar
and related jets are positionally coincident, and the small absorbing
column density observed for the X-ray spectrum (with an equivalent
$A_V \simeq 7$ mag, fully compatible with the absorbing column density
observed in the optical towards the jet) allow us to exclude that the
X-ray emission is associated with the protostellar sources (which are
hidden behind $\simeq 150$ mag of obscuration). The size of the jets
($\simeq 10$ arcsec) originating at L1551 IRS5 is smaller in angular
extent than the XMM EPIC PSF ($\simeq 14$ arcsec), so that no
inference is possible on spatial grounds about the possible detailed
location of the origin of the X-ray emission.

The emission from the IRS5 jet is compatible with being caused by
thermal emission from a plasma heated to a moderate temperature ($T
\simeq 0.5$~MK), equivalent to the shock temperature that is expected
at the interface (``working surface'') between the jet and the
surrounding circumstellar medium, on the basis of the observed jet
velocity. This is regardless of whether it is a bow-shock or a
reversed shock, which in itself is a function of whether the jet is
denser than the ambient medium or vice versa. \citet{fl98} find strong
evidence for the jet being less dense than the surrounding medium.
That conclusion was reached by estimating the mass in the jet from the
relative brightness of the shock, as well as discerning the ambient
density from molecular line measurements. Our observation of X-rays
from a $\simeq 0.5$~MK plasma is fully consistent with the conclusion
of \citet{fl98}. This is because the hypothesis of a ``light'' jet
requires a high degree of ionization (so that we visually observe all
the material present in the jet in recombination radiation). A low
degree of ionization at the observed shock velocities, would imply a
plasma temperature well above 1~MK, which is incompatible with our
X-ray data. We thus consider it most likely that the X-ray emission
originates directly from a shock associated with the so-called knot D,
and that our observations support the conclusion of \citet{fl98} that
the jet has a high degree of ionization.

The presence of a soft X-ray source at the position of the shock
associated with the jet is likely to have a significant influence on
the physical conditions of the accretion disk: while X-ray emission
from the star's corona is often likely to be more powerful that the
emission coming from the jet, most of the accretion disk (except for
the innermost part) is effectively shielded and not strongly
illuminated by the stellar coronal emission. The jet X-ray source on
the other hand lies above the disk (if indeed it is located at the
jet-circumstellar medium interface it is some 1000 AU above the disk),
illuminating the disk from above. This X-ray flux can therefore ionize
the disk material, and thus influence significantly the disk physical
conditions. Once more, if X-ray emission from protostellar jets is
indeed a common feature, this would influence the protostellar
environment significantly. This will be investigated in the future
through detailed modelling of the relative contribution of the stellar
coronal and jet X-ray luminosity to the ionization of the accretion
disk.

\begin{acknowledgements}
  
  GM, SS acknowledge the partial support of ASI and MURST. This paper
  is based on observations obtained with XMM-Newton, an ESA science
  mission with instruments and contributions directly funded by ESA
  Member States and the USA (NASA). Part of the data have been taken
  using ALFOSC, which is owned by the Instituto de Astrofisica de
  Andalucia (IAA) and operated at the Nordic Optical Telescope under
  agreement between IAA and the NBIfAFG of the Astronomical
  Observatory of Copenhagen. Nordic Optical Telescope is operated on
  the island of La Palma jointly by Denmark, Finland, Iceland, Norway,
  and Sweden, in the Spanish Observatorio del Roque de los Muchachos
  of the Instituto de Astrofisica de Canarias.

\end{acknowledgements}


\end{document}